\documentclass[journal=jacsat,manuscript=article,layout=traditional,super=false]{achemso}

\usepackage[version=3]{mhchem} 
\usepackage{graphicx} 
\usepackage{bm} 
\usepackage{amsmath,amssymb} 
\usepackage[T1]{fontenc}
\usepackage{textcomp, gensymb} 
\usepackage[labelfont=bf]{caption} 

\author{Jonathan G. Raybin}
\affiliation{Department of Chemistry, University of California, Berkeley, CA 94720, United States.}
\alsoaffiliation{Present address: National Insitute of Standards and Technology, Gaithersburg, MD, 20899, United States.}
\author{Ethan J. Dunsworth}
\affiliation{Department of Engineering Science, University of California, Berkeley, CA 94720, United States.}
\author{Veronica Guo}
\affiliation{Department of Physics, University of California, Los Angeles, CA 90024, United States.}
\alsoaffiliation{Present address: Department of Physics, Stanford, Stanford, CA, 94305}
\author{Naomi S. Ginsberg}
\email{nsginsberg@berkeley.edu}
\affiliation{Department of Chemistry, University of California, Berkeley, CA 94720, United States.}
\alsoaffiliation{Department of Physics, University of California, Berkeley, CA 94720, United States.}
\alsoaffiliation{Molecular Biophysics and Integrated Bioimaging Division, Lawrence Berkeley National Laboratory, Berkeley, CA 94720, United States.}
\alsoaffiliation{Materials Sciences \& Chemical Sciences Divisions, Lawrence Berkeley National Laboratory, Berkeley, CA 94720, United States}
\alsoaffiliation{Kavli Energy NanoScience Institute, Berkeley, CA 94720, United States.}
\alsoaffiliation{STROBE, NSF Science \& Technology Center, Berkeley, California 94720, United States.}

\title[An \textsf{achemso} demo]
 {Reversible Electron-Beam Patterning of Colloidal Nanoparticles at Fluid Interfaces}

\abbreviations{colloidal patterning, directed assembly, electron beams, interfaces, nanoparticles, SEM}

\begin{document}

\begin{tocentry}

\includegraphics[width=7.5 cm]{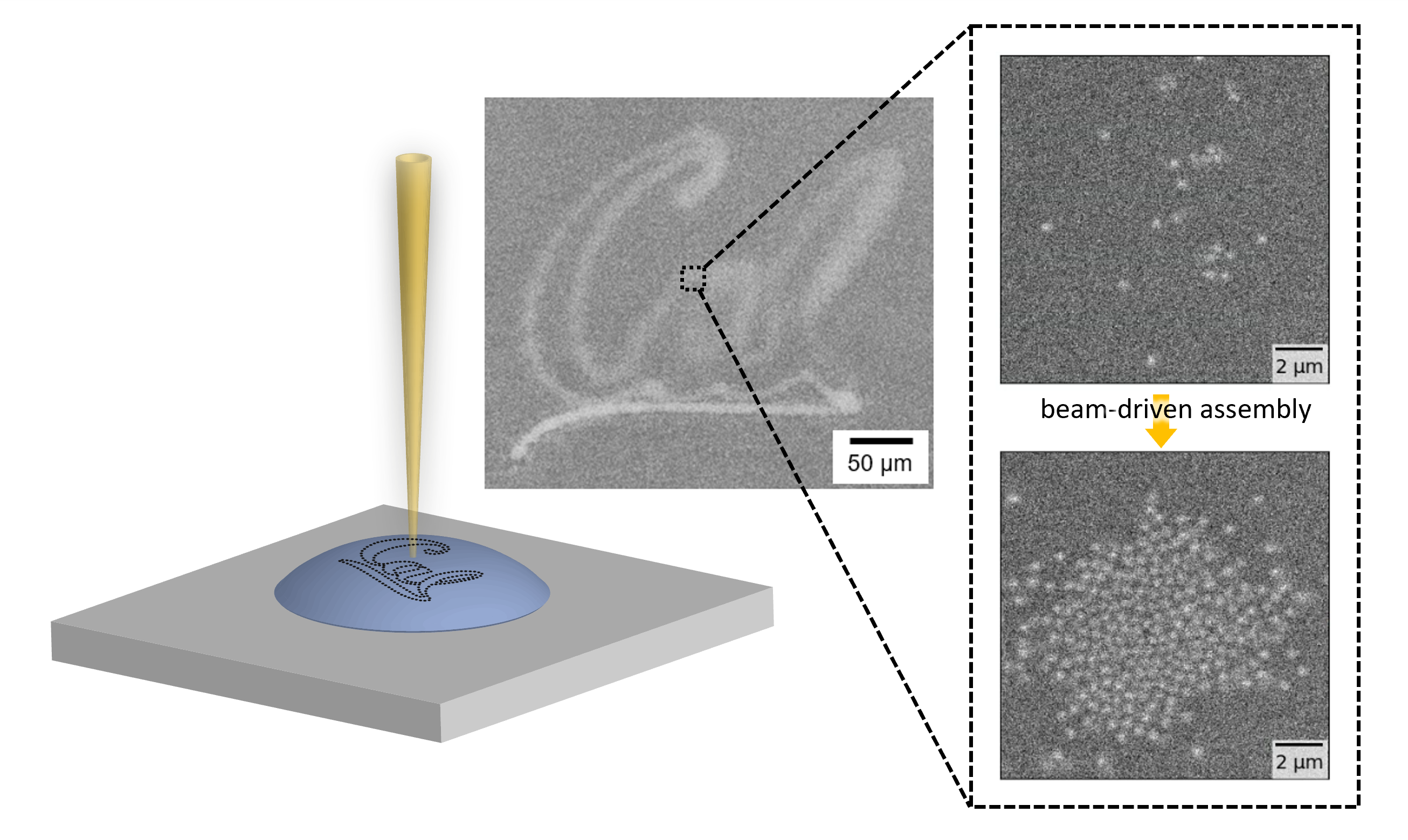}

\end{tocentry}

\begin{abstract}
The directed self-assembly of colloidal nanoparticles (NPs) using external fields guides the formation of sophisticated hierarchical materials but becomes less effective with decreasing particle size. As an alternative, electron-beam-driven assembly offers a potential avenue for targeted nanoscale manipulation, yet remains poorly controlled due to the variety and complexity of beam interaction mechanisms. Here, we investigate the beam-particle interaction of silica NPs pinned to the fluid-vacuum interface of ionic liquid droplets. In these experiments, scanning electron microscopy of the droplet surface resolves NP trajectories over space and time while simultaneously driving their reorganization. With this platform, we demonstrate the ability to direct particle transport and create transient, reversible colloidal patterns on the droplet surface. By tuning the beam voltage, we achieve precise control over both the strength and sign of the beam-particle interaction, with low voltages repelling particles and high voltages attracting them. This response stems from the formation of well-defined solvent flow fields generated from trace radiolysis of the ionic liquid, as determined through statistical analysis of single-particle trajectories under varying solvent composition. Altogether, electron-beam-guided assembly introduces a versatile strategy for nanoscale colloidal manipulation, offering new possibilities for the design of dynamic, reconfigurable systems with applications in adaptive photonics and catalysis.

\end{abstract}

\section{Introduction}

The self-assembly of colloidal nanoparticles (NPs) enables the creation of hierarchical materials with versatile structure and function. During self-assembly, the formation of desired structures may be encouraged by tuning various assembly parameters, including the NP core composition, surface chemistry, and solution environment.\cite{boles_self-assembly_2016} This flexibility has driven the development of a wide range of functional materials with applications in photonics, catalysis, energy storage, and chemical or thermal sensing.\cite{hou_patterned_2018, li_three_2021, hou_interfacial_2023} More sophisticated control over the assembly process may also be achieved through direct particle manipulation with external guiding fields.\cite{grzelczak_directed_2010, liljestrom_active_2019} By providing an external energy source, applied electric or magnetic fields access dissipative pathways, allowing for the formation of non-equilibrium or dynamic structures that extend beyond the scope of traditional assembly.\cite{sherman_dynamic_2016, harraq_field-induced_2022} Significant challenges, however, remain in scaling down this directed approach to finer length scales. As particle dimensions approach the nanoscale, the magnitude of dipolar interactions decreases and can fall below the thermal energy, causing dynamics to become dominated by Brownian motion.\cite{grzelczak_directed_2010} Additionally, fields have been typically applied on the laboratory scale where they act uniformly over the full particle ensemble, lacking targeted control.

Electron beam manipulation offers a compelling alternative strategy for directing NP assembly, taking advantage of electron microscopes’ ability to focus and align electron beams with sub-nanometer precision. This approach has been demonstrated for trapping or repelling individual particles\cite{zheng_electron_2012, white_charged_2012, chen_electron_2014}, guiding collective particle assembly\cite{liu_situ_2013}, and driving colloidal phase transformations.\cite{bischak_charging-driven_2020, raybin_nonadditive_2023} Still, across these examples, pinpointing specific beam-particle interactions proves challenging, as the interaction mechanism can vary significantly between experimental systems. As a result, reported beam effects are often poorly defined and provide limited control over the resulting colloidal response. This difficulty arises from the multitude of potential phenomena triggered by irradiating colloidal samples with a high-energy electron beam, including heating, charging, radiolysis, and solvent vaporization.\cite{schneider_electronwater_2014, yesibolati_unhindered_2020} Furthermore, most electron microscopy studies of fluid systems require sample encapsulation to prevent vacuum exposure, where particle adhesion to the membrane surfaces can hinder their dynamic response.\cite{yesibolati_unhindered_2020} In general, a comprehensive understanding of colloidal behavior requires disentangling the complex interplay of interactions between the beam, the NP system, and the environment. Measuring and elucidating the balance of each of these effects is therefore crucial for advancing electron beam manipulation as a reliable tool for guiding NP assembly.

Here, we investigate the beam-particle interaction of silica NPs confined to the fluid-vacuum interface of ionic liquid (IL) droplets. In these experiments, scanning electron microscopy (SEM) records particle trajectories while simultaneously driving their reorganization. The particles exhibit a well-defined and tunable response, which can be switched between attraction and repulsion by varying the beam voltage and dose. As a demonstration of this control, we use the scanning electron beam to write reversible and dynamic colloidal patterns at the droplet surface. To quantify the interaction strength, we employ single-particle tracking over many observations to statistically reconstruct the beam-induced force. This analysis reveals that the interaction is governed by induced hydrodynamic flow driven by trace radiolysis of the IL solvent. The role of IL radiolysis is further evidenced by experiments with varying solvent composition. Altogether, this improved understanding of the beam-particle interaction enables localized and dynamic control over NP organization, offering a new avenue for directed nanomaterial design.

\section{Results and Discussion}

Our experimental system consists of well-dispersed silica NPs (300, 500, or 1000 nm diameter) attached to the surface of a 1-ethyl-3-methylimidazolium ethyl sulfate ($\rm [EMIM]^+/[ETSO_4]^-$) droplet, as depicted in \textbf{Figure 1a}. The low vapor pressure of the IL solvent ensures compatibility with the SEM vacuum environment without encapsulation, enabling direct investigation of the fluid-vacuum interface.\cite{kim_visualizing_2016, kim_assessing_2019} Particles collect at the droplet surface to minimize surface tension, with each particle stabilizing the interfacial free energy by approximately $10^3-10^4$ $k_{\rm B}T$ where $k_{\rm B}$ is Boltzmann’s constant and $T$ is temperature.\cite{bischak_charging-driven_2020} The surface-bound particles consequently exhibit minimal height fluctuations, allowing us to treat the system as two-dimensional (2D). In these experiments, solutions were prepared with a low colloidal surface coverage fraction ($\phi < 0.01$) to minimize interparticle interactions and isolate the effects of beam exposure. At this concentration, particles have a mean separation of 2.5 $\mu \rm m$, significantly exceeding the approximately 10 nm electrostatic screening distance in IL solvent.\cite{lee_scaling_2017}

\textbf{\textit{Electron beam imaging and manipulation.}} In SEM images, particles appear as bright features over a dark IL background due to the increased electron scattering from silica (\textbf{Figure 1a}). Scattered electrons have a limited mean free path in the IL solution, such that imaging is surface-sensitive and resolves particles with depths of no more than 500 nm, depending on the electron beam voltage.\cite{kim_visualizing_2016} Our imaging therefore selectively tracks the in-plane dynamics of particles pinned to the IL-vacuum interface. Movies are acquired through repeated raster scanning, with frame rates of 0.8-3.3 Hz. Across this range, the scanning rate of the beam is much faster than particle dynamics, allowing us to unambiguously track their trajectories between frames.

\begin{figure*}
\includegraphics[width=6.4 in]{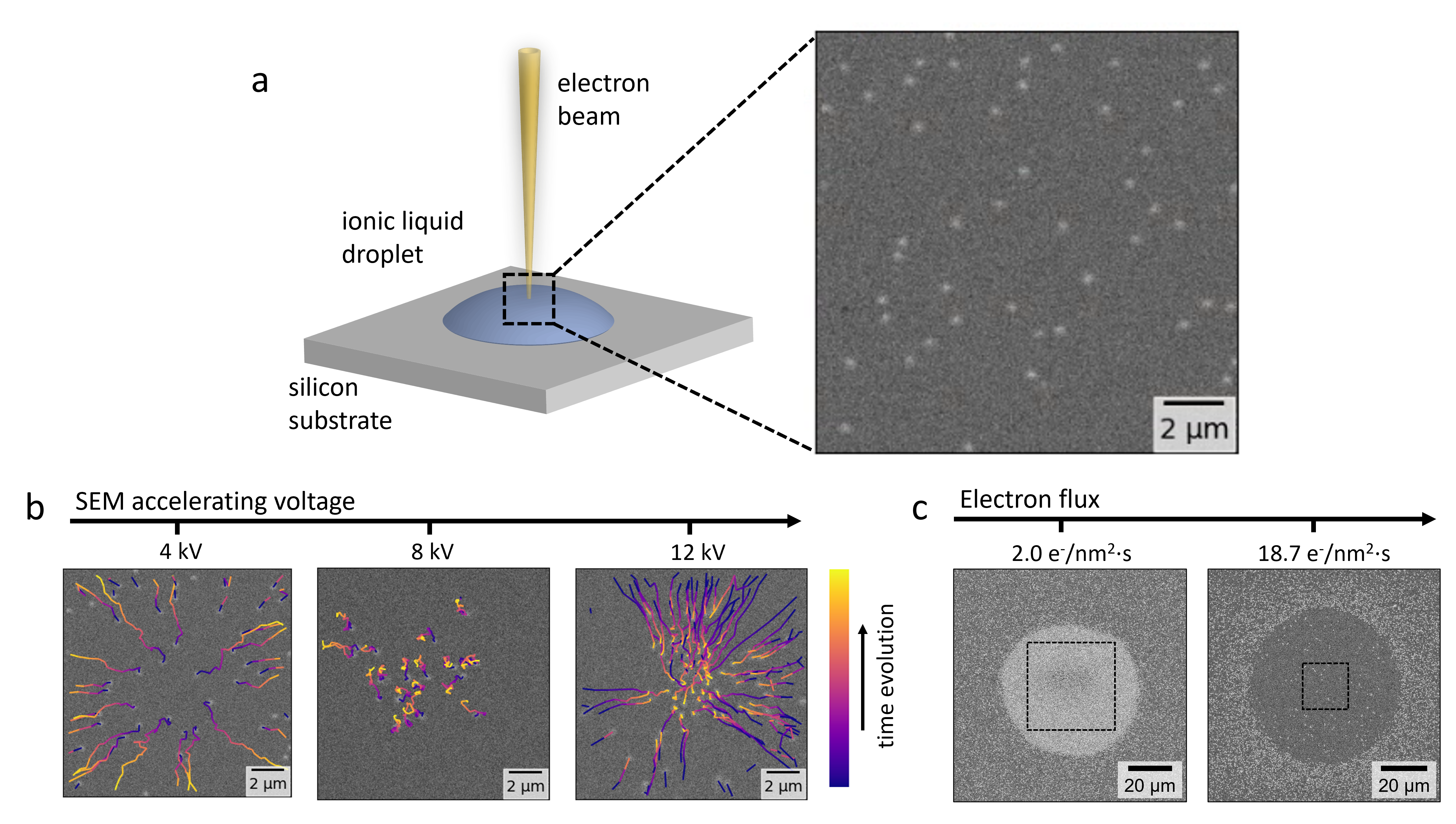}
\caption{\textbf{Overview of particle imaging and manipulation.} (a) Experimental schematic and SEM imaging of dispersed 500-nm-diameter silica NPs at the IL droplet interface. (b) The particle response with varying accelerating voltage and constant electron dose (0.37 $e^-/ \rm (nm^2 \cdot s)$). Particles are repelled from the scan area at low voltage (4 kV), exhibit approximately Brownian motion at intermediate voltage (8 kV), and are attracted at high voltage (12 kV). The time evolution of single-particle trajectories (from purple to yellow) is depicted over a 42 s (32 frame) period. (c) At constant voltage (12 kV), particles are accumulated at low dose (2.0 $e^-/ \rm (nm^2 \cdot s)$) and depleted at high dose (18.7 $e^-/ \rm (nm^2 \cdot s)$). Images show zoomed-out views of the particle density following repeated raster scanning over the area indicated by the dashed box.
}
\label{F1}
\end{figure*}

In addition to resolving particle positions over space and time, SEM imaging also elicits a dynamic response. The response is robust across imaging systems, as evidenced by consistent observations of colloidal dynamics using two SEMs with distinct imaging characteristics. Interestingly, under low-electron-dose conditions ($< 5.0$ $e^-/ \rm (nm^2 \cdot s)$), we found that varying the beam voltage controls both the strength and direction of the particle interaction. Particles are repelled from the center of the imaging area at low voltages and are attracted to it at high voltages, switching between these regimes at approximately 8 kV. This tunability contrasts with previous examples of electron-beam-particle interactions, which found either uniform attraction\cite{zheng_electron_2012, liu_situ_2013, chen_electron_2014} or uniform repulsion.\cite{white_charged_2012, raybin_nonadditive_2023}

To visualize the response, we followed individual particle trajectories, as shown for 500-nm-diameter NPs in \textbf{Figure 1b} and \textbf{Supporting Videos S1-S3}. The color scale indicates the particle time evolution (from purple to yellow), showing the bias for outward motion at 4 kV and inward motion at 12 kV. At the 8 kV crossover voltage, the beam interaction is minimized, and particles exhibit approximately Brownian trajectories.

The response is also influenced by varying the electron dose.
Below 5.0 $e^-/ \rm (nm^2 \cdot s)$, tuning the electron flux controls the NP interaction strength without affecting its sign. At higher doses ($>10.0$ $e^-/ \rm (nm^2 \cdot s)$), however, the beam interaction increases to the point that it overcomes surface pinning forces, and particles are able to detach from the droplet interface. In this regime, SEM movies show continuous and rapid particle circulation with dynamics too fast for reliable particle tracking. This difference is evident in zoomed-out perspectives of the particle coverage following repeated raster scanning at low and high dosage in the same droplet sample at 12 kV (\textbf{Figure 1c}). SEM movies similarly show stark differences in the particle response (\textbf{Supporting Videos S4 and S5}). With accumulated exposure at low dose, particles are attracted to the image area, where they collect into a dense (light contrast) raft. As the particles crowd together, their dynamics become kinetically arrested, and they jam to form a disordered packing. At high dose, however, particles detach from the interface, leading to local depletion of the exposed (dark contrast) region. In each case, the effects of beam irradiation extend over a larger area than the imaging field of view (FOV), delineated by the dashed boxes in \textbf{Figure 1c}, yielding an approximately circular, axisymmetric profile.

\textbf{\textit{Reversible assembly of colloidal patterns.}} The ability to control local particle density suggests a natural method for patterning larger-scale colloidal assemblies. By sequentially exposing specified regions with the electron beam, we can organize colloidal NPs to achieve a variety of target structures (\textbf{Figure 2a-b}). Additionally, the tunability of the beam interaction enables either particle accumulation or depletion for the preparation of positive or negative patterns, respectively. As an example, the same programmed exposure (\textbf{Figure 2a(i)}) generates a positive pattern using a low electron flux (2.0 $e^-/ \rm (nm^2 \cdot s)$, \textbf{Figure 2a(ii)}) and a negative pattern at a high electron flux (18.7 $e^-/ \rm (nm^2 \cdot s)$, \textbf{Figure 2a(iii)}). A more detailed discussion of the pattern-writing process is included as Supporting Information (\textbf{Figure S1}). At this magnification, SEM imaging no longer resolves individual particles, but variation in the local particle density is apparent from changes in the overall electron scattering contrast. To improve the contrast, these patterns were formed using a more concentrated colloidal solution than used for the single-particle tracking experiments, with a surface coverage fraction of $\phi \sim 0.03$.

\begin{figure*}
\includegraphics[width=4.8 in]{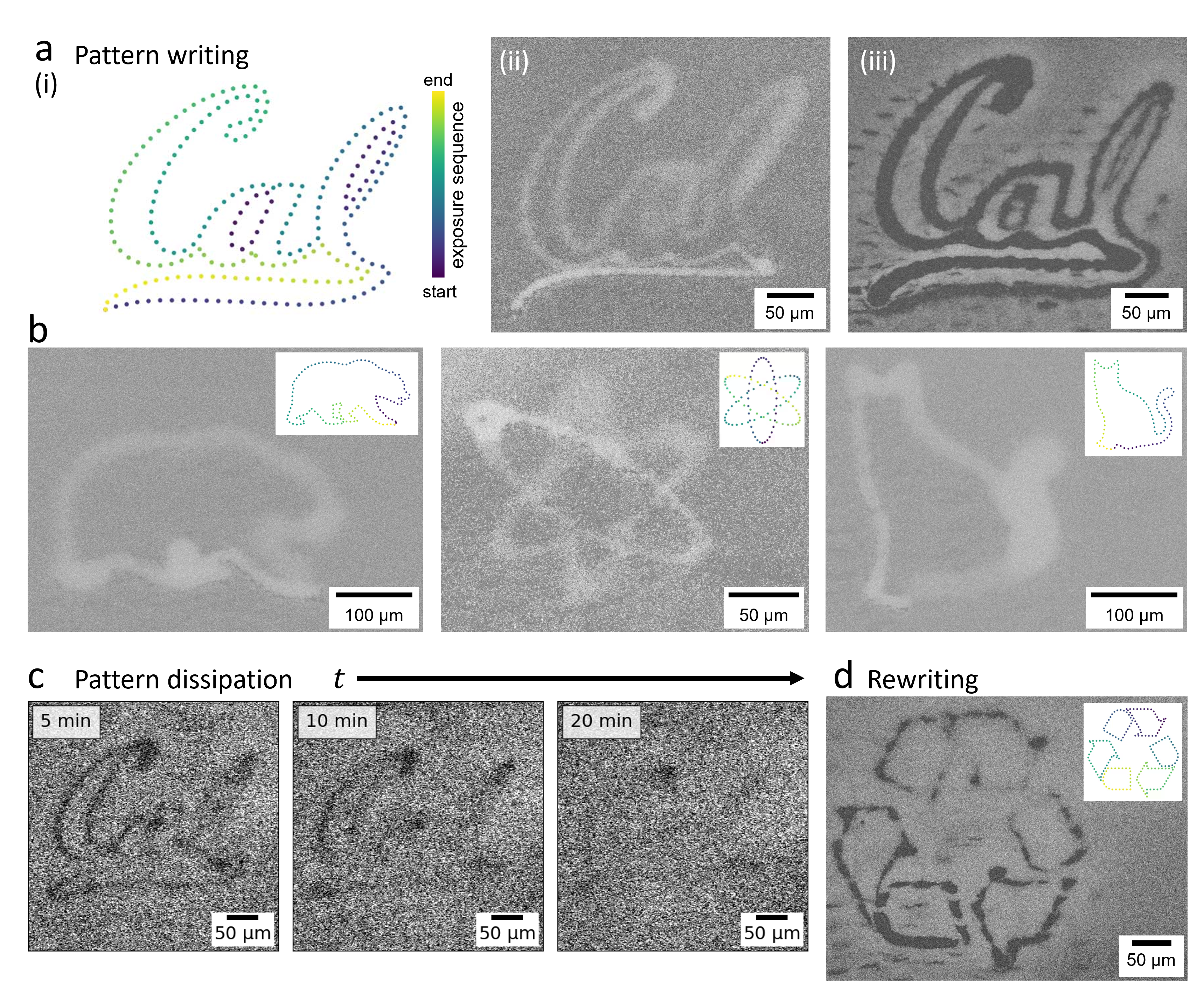}
\caption{\textbf{Reversible pattern writing and dissipation.} (a) Electron beam exposure sequences (i) generate corresponding colloidal patterns at the droplet interface. Positive patterns (ii) are formed using low electron dose (2.0 $e^-/ \rm (nm^2 \cdot s)$), and negative patterns (iii) are formed with high electron dose (18.7 $e^-/ \rm (nm^2 \cdot s)$). (b) Changing the exposure sequence (inset) offers flexible control over the resulting pattern (each formed at the same low electron dose as in (a-ii)).  (c) The patterns remain fluid and dissipate over time. (d) New patterns may be subsequently rewritten at the same location (shown for high electron dose as in (a-iii)).
}
\label{F2}
\end{figure*}

This pattern-writing procedure, which leverages the nanoscale focus and precision of electron microscopy, is closely analogous to electron beam lithography except that it is reversible and occurs on a fluid bed. It therefore offers many of the same advantages, including flexible control over the pattern scale and shape. Unlike traditional lithographic patterning of a solid resist material, however, in IL droplets the electron beam generates flow fields that sculpt the particle distribution at the liquid surface. Because the patterns remain fluid, they intrinsically dissipate over time through the diffusion of their constituent colloidal NPs (\textbf{Figure 2b}; \textbf{Supporting Video S6}). Eventually the surface recovers the initial state with uniform colloidal coverage, at which point a new pattern can be written. An example of this rewritability is shown in \textbf{Figure 2c}, where the new pattern was drawn over the same location as \textbf{Figure 2b}. This process may be repeatedly iterated -- akin to a microscale ``Etch A Sketch” -- and we were able to work continuously with the same droplet sample over periods of at least 4 hours with no apparent change in the particle response.

The ability to dynamically reconfigure colloidal solutions is relevant to many different fields. In e-ink displays, for example, contrast is controlled using modulated electrostatic interactions to shuttle colloidal pigments. Similarly, in photonics, systems of responsive particles have been shown to transition to a random lasing state when the local scatterer density exceeds a critical threshold for optical amplification.\cite{trivedi_self-organized_2022} In heterogeneous catalysis, adjusting the density of catalytic particles modulates their activity, enabling the creation of adaptive microreactors capable of maintaining chemostasis\cite{milster_synergistic_2023} or driving oscillatory reactions.\cite{howlett_autonomously_2022} These examples, along with the fundamental goal of preparing structured fluid systems,\cite{moller_writing_2023} highlight the demand for precise and flexible control over colloidal assembly.

\textbf{\textit{Statistical analysis of the beam-particle interaction force.}} In order to explain the particle response, we apply single-particle tracking to measure the effective beam-particle interaction force in the low-dose regime. Alongside the beam interaction, colloidal particles are subject to Brownian motion, causing each individual track to follow a stochastic trajectory along the droplet surface. We therefore perform a statistical analysis over an ensemble of many such trajectories (\textbf{Figure 3a}) to reconstruct the forces underlying their dynamics. The resulting ensemble probability density $W(\boldsymbol{q},t)$ evolves in time following the Fokker-Planck equation:
\begin{equation}
{
\frac{\partial W(\boldsymbol{q},t)}{\partial t}
= L_{\rm FP}(\boldsymbol{q},t)W(\boldsymbol{q},t), 
}
\label{EQ1}
\end{equation}
where $t$ is time and the vector $\boldsymbol{q}$ represents the set of observable system parameters, in this case the 2D particle positions.\cite{beltran-villegas_fokkerplanck_2010} In this expression, the Fokker-Planck operator $L_{\rm FP}(\boldsymbol{q},t)$ is defined:
\begin{equation}
{
L_{\rm FP}(\boldsymbol{q},t)
= \sum_i \frac{\partial}{\partial q_i} D^{(1)}_i(\boldsymbol{q},t)
+ \sum_{i, j} \frac{\partial^2}{\partial q_i \partial q_j} D^{(2)}_{ij}(\boldsymbol{q},t).
}
\label{EQ2}
\end{equation}
The operator is specified by drift and diffusion coefficients $D^{(1)}_i$ and $D^{(2)}_{ij}$, with $i$ and $j$ indexing the degrees of freedom of $\boldsymbol{q}$. In the present case, where $\boldsymbol{q}$ corresponds to spatial coordinates, this expression is a generalization of the Smoluchowksi diffusion equation:
\begin{equation}
{
\frac{\partial W(\boldsymbol{q},t)}{\partial t} 
= \nabla \cdot \boldsymbol{D} \cdot \left( \nabla - \frac{\boldsymbol{F}}{k_{\rm B}T} \right) W(\boldsymbol{q},t),
}
\label{EQ3}
\end{equation}
where $\boldsymbol{D}$ is diffusion and $\boldsymbol{F}$ is the external force.\cite{chavanis_generalized_2019} Because Fokker-Planck analysis involves minimal physical assumptions (solely that the process be continuous and Markovian), it provides a suitably agnostic measure of the interaction force independent of its specific mechanism.

Within this framework, identification of the drift and diffusion coefficients is sufficient to fully specify the evolution of all ensemble properties of the system. Assuming steady-state, these coefficients may be determined from the mean and variance, respectively, of the frame-to-frame displacement vectors $\delta \boldsymbol{q}$.\cite{beltran-villegas_fokkerplanck_2010} Because the trajectory ensemble is radially symmetric, we express these vectors in polar coordinates with radial and azimuthal components $\delta q_r$ and $\delta q_{\theta}$ (\textbf{Figure 3b}). To assess spatial variation, we also divide the displacement data into concentric radial bins, as indicated in \textbf{Figure 3a}. 

\begin{figure*}
\includegraphics[width=6.4 in]{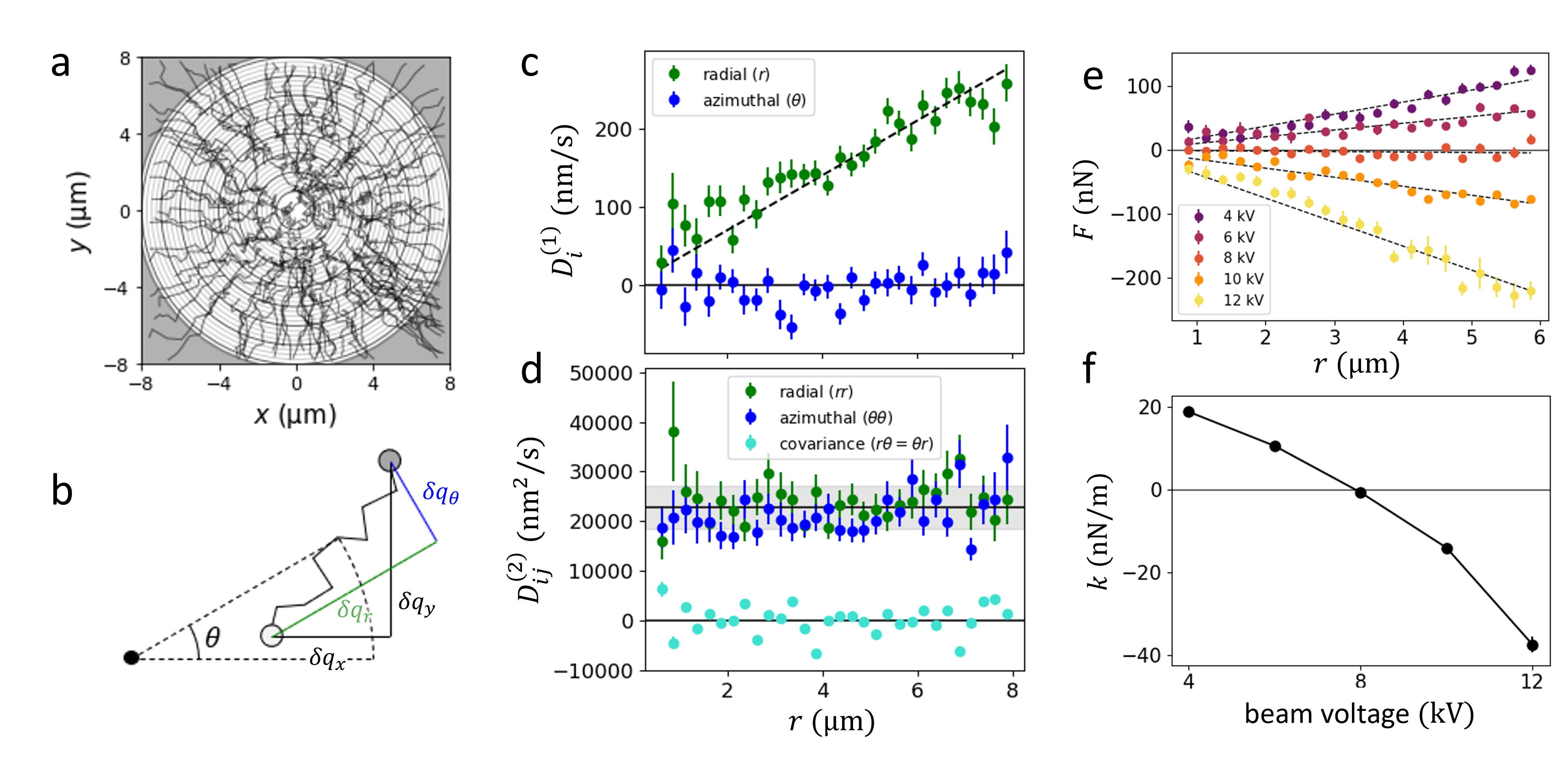}
\caption{\textbf{Fokker-Planck analysis of single-particle trajectories.} (a) Single-particle trajectories of 300-nm-diameter particles are collected over 13 trials and (b) transformed to polar coordinates. Measurement of the (c) drift and (d) diffusion coefficients from statistical analysis of the frame-to-frame displacement vectors $\delta \boldsymbol{q}$ reveals uniform diffusivity and a radially aligned interaction force. (e) The magnitude of beam-particle interaction force increases linearly from the image center (shown for 500-nm-diameter particles), and (f) the effective interaction strength can be tuned from repulsive ($k > 0$) to attractive ($k < 0$) by varying the electron beam voltage.
}
\label{F3}
\end{figure*}

Our analysis of the drift term (\textbf{Figure 3c}) shows that the beam-particle interaction is oriented radially and increases with distance from the center of the FOV. The radial drift component $D^{(1)}_r$ (green) increases linearly from the center, while the azimuthal component $D^{(1)}_{\theta}$ (blue) remains zero across the full image area. Meanwhile, the diffusive term is constant over the FOV, with a value of $D=2.3(4) \times 10^{-10}$ $\rm cm^2/s$, consistent with estimates obtained from the Stokes-Einstein relation for diffusion in IL.\cite{requejo_effect_2014} We emphasize that this uniform diffusivity is not a prior assumption of our model. In our analysis of the diffusive term (\textbf{Figure 3d}), the diagonal matrix elements $D^{(2)}_{rr}$ (green) and $D^{(2)}_{\theta \theta}$ (blue) are constant and equal, indicating isotropic and uniform 2D diffusion within the imaging area. The off-diagonal terms $D^{(2)}_{r \theta} = D^{(2)}_{\theta r}$ (cyan) are zero, which confirms dimensional independence. Altogether, these results affirm that the diffusion tensor may be represented by a single scalar constant. This uniformity indicates that electron beam irradiation does not significantly disrupt particle diffusion, for instance by altering the local solvent viscosity.

Comparing Equation (2) with the Smoluchowski equation (3), the interaction force simplifies to $F=\frac{k_{\rm B} T}{D} \boldsymbol{D}^{(1)}$ in the case of uniform diffusion. From our drift measurement, the force profile increases linearly from zero at the image center following $F(\boldsymbol{q})=kq_r$, with $k=8.4(3) \times 10^{-9}$ N/m. The parameter $k$ serves as a negative effective spring constant that describes the strength of the repulsive interaction. For comparison, this force gradient is much weaker than the trapping forces involved in optical tweezers, which have typical interaction strengths of order $10^{-6}$ N/m.\cite{polimeno_optical_2018} Across the image area, the driving force is modest compared to particle diffusion, yielding P\'{e}clet  numbers close to 1. Toward the center of image, where the interaction force is weakest, $Pe < 1$ and particle transport is primarily driven by Brownian motion (\textbf{Figure S2)}.

Varying the beam voltage provides continuous and tunable control over the measured interaction strength. Repeating this analysis across the full voltage range (4-12 kV) using 500-nm-diameter NPs, we observe consistent transport behavior characterized by uniform diffusivity and radially oriented drift (\textbf{Figure S3}). The magnitude of the interaction force exhibits a similar radial profile at each voltage, increasing linearly from the center of the FOV (\textbf{Figure 3e}). With increasing voltage, the interaction strength k steadily decreases and reverses sign from positive to negative at a crossover voltage of approximately 8 kV (\textbf{Figure 3f}). This tunability is maintained across different particle sizes, showing an equivalent trend for 300-nm NPs (\textbf{Figure S4a}).

Taken together, these measurements of the beam interaction force offer a surprisingly simple result. During image capture, the scanning beam sweeps a continuous raster over the full FOV, suggesting, at least in principle, a more complicated interaction dependent on the moving beam trajectory. Yet, because the motion of the beam is much faster than the particle response, the measured interaction force represents an average taken over the imaging time scale. From this perspective, the effective exposure pattern corresponds to an approximately uniform irradiation pattern based on the superposition of multiple beam positions over time. Our observation of radially symmetric particle trajectories supports this simplified picture, where the effective interaction force is entirely specified by each particle’s relative location within the scan area.

\textbf{\textit{Beam-induced flow drives particle transport.}} Based on the observed radial interaction profile, we developed a model to explain the colloidal transport mechanism. We propose that the effective beam-particle interaction results from solvent flow induced by trace radiolysis of the IL ions, as depicted in \textbf{Figure 4a}. During electron beam exposure, injected electrons are captured by imidazolium cations to generate a dilute concentration of neutral $\rm \pi$-radicals.\cite{kim_visualizing_2016} In solution, these radicals engage in a cascade of secondary reaction pathways, which include coupling reactions with other IL cations to form resonance-stabilized dimers and oligomers.\cite{shkrob_deprotonation_2010, mincher_radiation_2014, rola_interaction_2021} In our model, the steady-state enrichment of cross-linked IL species leads to uniform downward flow of solvent within the imaging FOV. By continuity, fresh solvent is drawn into the imaging area along the droplet surface, circulating in a steady-state toroidal flow pattern. Throughout this process, particles are entrained by solvent currents while remaining pinned to the droplet interface, causing them to collect at the center of the recirculating flow field.

\begin{figure*}
\includegraphics[width=6.4 in]{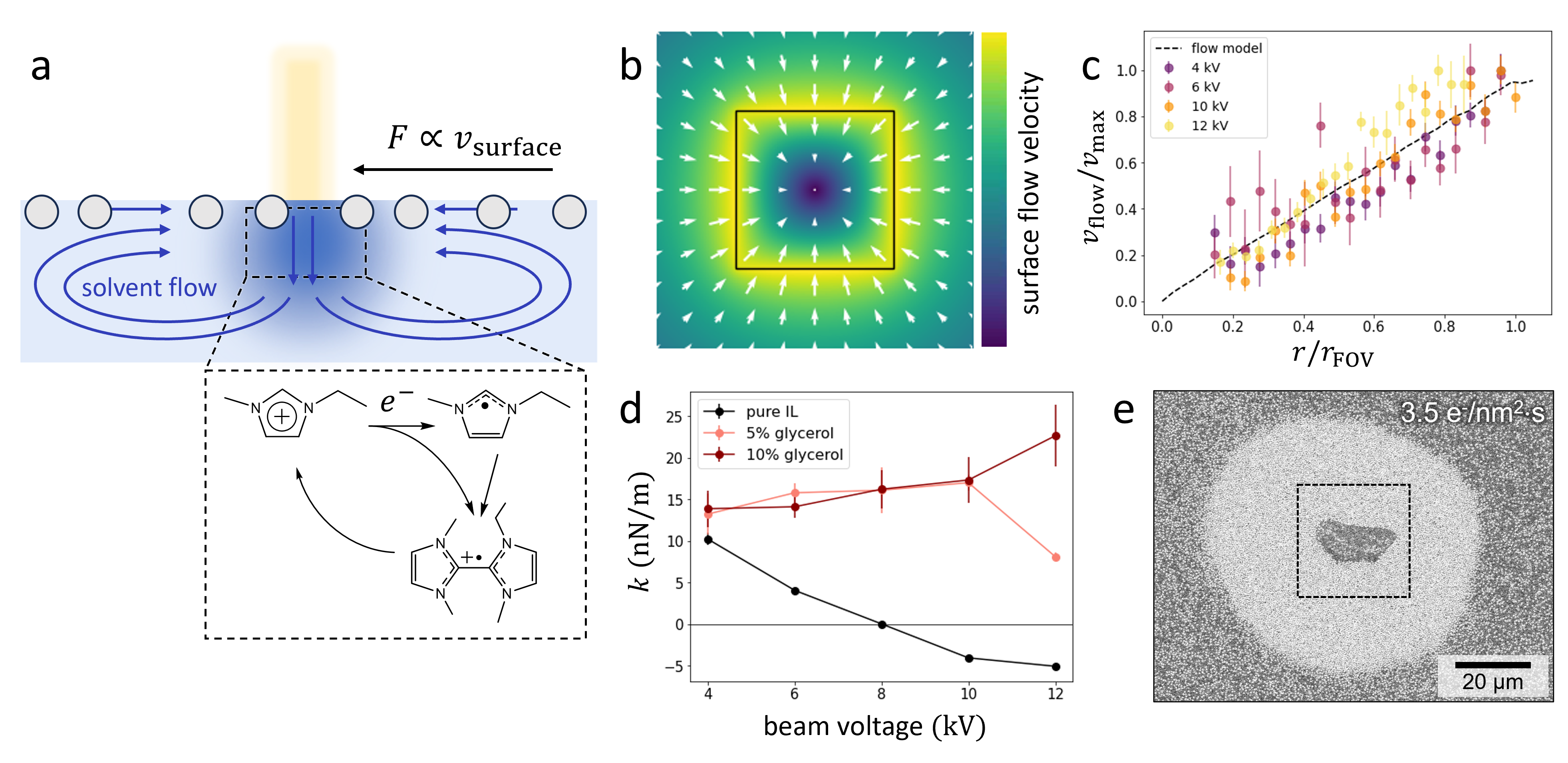}
\caption{\textbf{Overview of the beam-particle interaction mechanism.} (a) Schematic illustrating the beam-induced transport mechanism. Electron beam irradiation generates cross-linked imidazolium species through radical coupling reactions. Sinking of cross-linked material from the irradiated volume (dark blue shaded region) drives the formation of steady-state solvent currents (blue arrows). Surface-bound colloidal particles are entrained by the near-surface flow leading to an effective beam-particle attraction (black arrow). (b) Modeled surface flow from uniform electron-beam exposure over a square raster area (black box) shows a nearly symmetric radial profile. (c) The modeled flow velocity increases linearly from the center, consistent with experimental particle velocities. (d) The addition of glycerol suppresses colloidal transport by mitigating IL radiolysis (shown for 300-nm-diamter particles). In pure IL (black), the interaction strength $k$ is voltage-dependent and switches from repulsive to attractive, while in 5\% (pink) or 10\% (red) glycerol the interaction is consistently repulsive. (e) Formation of a cross-linked IL gel (dark region) surrounded by a particle raft at intermediate electron flux (3.5 $e^-/ \rm (nm^2 \cdot s)$). The image shows a zoomed-out view following repeated raster scanning over the indicated area (dashed box).
}
\label{F4}
\end{figure*}

We evaluated the validity of this model in two steps: first, by establishing that the particle response follows the beam-induced solvent flow pattern along the droplet surface, and, second, by confirming that the flow is driven by IL radiolysis. To describe the beam interaction, we conducted a numerical simulation of the solvent flow velocity along the droplet surface. In this 2D description -- focusing on solvent transport at the interface -- out-of-plane flow is represented as a distribution of sources or sinks. Here, we assumed uniform electron irradiation over the imaging area, resulting in a sink distribution that is constant within the square raster area and zero elsewhere. We neglected the recirculating upward flow, which is distributed over a much larger volume that extends radially well beyond the scan area. The in-plane flow velocity was then obtained by numerically solving the continuity equation in and around the scan area. Complete details of this analysis are included as Supporting Information (\textbf{Figure S5}). The resulting steady-state flow velocity field (\textbf{Figure 4b}) exhibits nearly axisymmetric inward flow, despite the square shape of the exposure pattern (indicated by the black box). Deviations from radial symmetry are only apparent near the corners of the scan area, as discussed in the Supporting Information. This symmetry causes flow to converge at the center of the sink distribution, creating a stagnation point. Azimuthal integration of the velocity field reveals that the magnitude of the flow velocity increases linearly as a function of radial distance from the center within the image area, agreeing with the experimentally measured particle velocities (\textbf{Figure 4c}). 

Our solvent flow model successfully captures both the symmetry and the linear force profile of particles bound to the IL-vacuum interface. The model also explains the accumulation of particles at the central stagnation point. Furthermore, it predicts that inward flow extends beyond the image area, consistent with our observations of long-range particle capture (\textbf{Figure 1c}). Outside the exposed region, the flow velocity decays as $1/r$ (\textbf{Figure S5}), and advective trapping forces are expected to continue to exceed Brownian motion out to distances of roughly 2-3 times the FOV radius.

Colloidal transport through solvent advection is a widespread phenomenon, observed in diverse scenarios such as the formation of coffee rings\cite{deegan_capillary_1997}, the clustering of Cheerios\cite{vella_cheerios_2005}, and the aggregation of NPs during solvent evaporation.\cite{rabani_drying-mediated_2003} Our colloidal patterning methodology similarly leverages solvent currents, using the electron beam to generate well-defined flow fields that guide NP organization. This targeted approach is analogous to the dissipative assembly procedure developed by Makey \emph{et al}.,\cite{makey_universality_2020} which uses ultrafast laser pulses to generate solvent currents that drive colloidal assembly.

Under this description, the measured particle interaction force $F(\boldsymbol{q})$ corresponds to hydrodynamic drag and is expected to be proportional to the solvent flow velocity, independent of particle size. To verify this prediction, we further examined colloidal solutions containing a mixture of 300- and 1000-nm-diameter particles. We found that differently sized particles responded to the electron beam in tandem, traveling together with the same velocity (\textbf{Supporting Video S7}). This behavior confirms that the particles serve as tracers of the underlying solvent flow field.

Having established the role of solvent flow, we next consider its origin. In principle, electron-beam-induced flow may arise through a number of possible mechanisms; According to the Onsager reciprocal relations, gradients in any thermodynamic quantity (including charge, heat, and ion concentration) lead to corresponding flow.\cite{onsager_reciprocal_1931} To test the specific role of radiolysis chemistry, we examined the beam-particle response in solutions containing dilute quantities of glycerol, an efficient radical scavenger.\cite{merenda_validation_2008} We found that adding glycerol effectively suppresses the voltage-dependence of the particle interaction (\textbf{Figure 4d}). In pure IL solutions (black), the particle response shows a strong voltage dependence, with the effective spring constant k switching from repulsive to attractive. By contrast, the 5\% (pink) and 10\% (red) glycerol solutions both show consistent particle repulsion, with differences in the response only appearing at 12 kV.

If flow were governed by direct heating or charging, we would expect minimal difference in NP behavior in the pure IL and glycerol solutions. Instead, the change may be understood through the context of our proposed radiolysis mechanism. At low voltages, particles in each solution are repelled from the image area. We attribute this baseline repulsion to electrostatic or electrophoretic interactions between the charge injected into the solution and the charged silica particles.\cite{white_charged_2012} As the beam voltage is increased, the higher energy electrons multiply-scatter over an expanded interaction volume, facilitating radical formation and solvent cross-linking.\cite{minamimoto_polymerization_2015, kim_visualizing_2016} In pure IL, induced solvent flow from the sinking of cross-linked species counteracts the baseline repulsion, enabling a voltage-tunable response. With added glycerol, however, radical capture competes with cross-linking reactions of the imidazolium, preventing beam-induced flow. For most accelerating voltages, glycerol is present in sufficient excess relative to the trace radicals generated, such that both 5\% and 10\% glycerol completely suppress the cross-linking reaction. At 12 kV, differences emerge between the glycerol solutions, as inward flow is only partially hindered at 5\% glycerol and is fully prevented at 10\% glycerol.

As additional evidence of the role of solvent radiolysis, SEM imaging directly captures the accumulation of cross-linked IL. At intermediate electron flux (3.5 $e^-/ \rm (nm^2 \cdot s)$), cross-linked material builds up over time to form a dark patch, inferred to be a solidified gel (\textbf{Figure 4e}). Surrounding particles are excluded from the solidified region, which adopts a roughly rectangular or trapezoidal shape due to the square SEM raster pattern. Particles embedded in the gel remain fixed, even as surrounding particles continue to exhibit Brownian motion. We do not observe gel formation under the low-electron-flux conditions (below 2.0 $e^-/ \rm (nm^2 \cdot s)$) used for pattern writing and single-particle tracking, as solvent circulation continuously replenishes the irradiated region and prevents the accumulation of cross-linked species. As particles fill the scan area, the colloidal monolayer serves as a protective barrier that shields the IL from direct electron exposure and slows the rates of both particle transport and continued gel formation. These observations highlight the interrelatedness of IL cross-linking and solvent flow, both originating from a common radiolysis mechanism.

Our understanding of the interaction mechanism has direct implications for the use of electron beams to trap and assemble particles on a liquid bed. Typically, trapping particles with solvent flow requires the formation of vortices due to the inherent continuity of hydrodynamic transport.\cite{li_three_2021} Irrotational and incompressible flow fields have zero divergence and, in analogy to Earnshaw’s theorem for electrostatics, cannot support stationary points for particle trapping. These constraints, however, no longer apply when particles are pinned to a droplet interface. In this scenario, particles follow only the in-plane component of the flow, which is able to converge, as shown in \textbf{Figure 4b}. Therefore, our ability to collect particles with the electron beam is achieved through the combined influence of both solvent flow and surface confinement. This mechanism also defines the limits of particle trapping. In the high-dose regime, solvent flow intensifies to the point that hydrodynamic friction exceeds surface confinement forces, causing particles to detach from the interface. Instead of being trapped, the particles rapidly circulate into the bulk droplet (\textbf{Supporting Video S5}).

The interaction mechanism additionally contributes to the observed reversibility of colloidal pattern writing depicted in \textbf{Figures 2c-d}. Although assembly is fueled by irreversible cross-linking of the IL solvent, only a trace proportion of ions is consumed through the pattern-writing process. The electron beam interaction volume is of order $10^{-9}$ nL, which represents a negligible fraction of the total 300 nL droplet volume.\cite{kim_visualizing_2016} Sinking of the cross-linked species induces solvent circulation, continuously replenishing exposed regions with fresh solvent from the IL reservoir. This mechanism ensures that patterns can be repeatedly erased and rewritten on the droplet surface over extended periods.

\section{Conclusion}
In summary, we have investigated the interaction of a scanning electron beam with colloidal particles confined to a fluid-vacuum interface. We find that the particles exhibit a tunable response with changing electron beam voltage and dose, offering the unique ability to switch between attractive and repulsive regimes. Through statistical analysis of particle trajectories, we separate the contributions of diffusion and drift to quantify the spatially varying interaction force with piconewton sensitivity. By varying solvent chemistry, we further deduce that the effective interaction results from flow driven by trace radiolysis of the IL solvent. With this improved understanding of the interaction forces and mechanism, we demonstrate the ability to use electron beams for assembling and patterning colloids at a droplet interface.

Looking forward, our approach leverages the nanoscale precision of SEMs to provide new opportunities for manipulating colloidal assembly at smaller length scales. A key advantage of this platform lies in the utility of electron microscopy for simultaneously imaging and manipulating NP systems. The integration of real-time feedback from SEM imaging could therefore offer dynamic control over the colloidal organization, facilitating the preparation of defect-free lattices or non-equilibrium assemblies.\cite{juarez_feedback_2012, tang_optimal_2016} The dynamically tunable interaction could also be harnessed for the development of switchable photonic or catalytic systems that can be reconfigured on demand.\cite{solomon_tools_2018}

Furthermore, the role of the IL solvent in mediating the particle response suggests potential avenues for further refining particle manipulation by varying the IL composition. Although the cross-linking mechanism examined here is specific to imidazolium-based ILs, similar polymerization chemistry has been reported for ILs containing ammonium cations.\cite{rola_interaction_2021} More broadly, ILs serve as ``designer solvents,” where each ionic component can be independently tuned to support specific radiation chemistry or enhance the electron scattering cross section.\cite{feldmann_ionic_2017} Mapping NP interactions across this space could reveal new strategies for optimizing electron-beam-driven assembly processes. For example, controlled cross-linking of the IL solvent during particle assembly could enable the preparation of functional nanocomposite materials that preserve the colloidal organization. Ultimately, an improved fundamental understanding of electron beam interactions -- with both NPs and solvents -- will pave the way for more sophisticated and tailored approaches to colloidal assembly at the nanoscale.

\section{Methods}
\subsection{Colloidal solutions}
Colloidal NPs were collected by centrifuging stock aqueous solution of silica nanospheres with bare silanol surface chemistry (285 $\pm$ 18, 518 $\pm$ 20, or 1030 $\pm$ 22 nm diameter, 10 mg $\rm mL^{-1}$, nanocomposix). The particles were then redispersed in 1-ethyl-3-methylimidazolium ethyl sulfate ($\rm [EMIM]^+/[ETSO_4]^-$) ionic liquid at a concentration of 50 mg $\rm mL^{-1}$. The solutions were heated overnight in a vacuum oven at 60 $\rm \deg C$ and ~20 torr to remove excess water. For glycerol tests, an additional 5 or 10 vol.\% glycerol was added to the dried solution. Next, 300 nL droplets were deposited onto cleaned $\sim 1$ cm $\times$ 1 cm Si wafer substrates (Virginia Semiconductor). The substrates were cleaned by solvent rinses with isopropyl alcohol, acetone, and distilled water followed by 2 min of oxygen plasma cleaning. The droplet samples were then loaded into the SEM chamber immediately following deposition to minimize air exposure.

\subsection{Scanning electron microscopy}
Electron microscopy was performed using two SEMs: a Gemini SUPRA 55 S2 SEM (Zeiss) and a Phenom Pharos G2 Desktop FEG-SEM (ThermoFisher Scientific). The particle response was independent of the imaging system and showed identical behavior under the same imaging conditions (beam voltage, beam current, magnification), although the Pharos was limited to beam voltages above 10 kV and could not access the repulsive interaction regime observed below 8 kV. Movies of the beam-particle interaction and single-particle tracking data were collected using the SUPRA. The examples of controlled particle assembly and pattern writing were performed using the Pharos.

\subsection{Single-particle tracking and analysis}
SEM images were analyzed using custom python code. In each frame, we applied a Laplacian of Gaussians filter to detect particle positions by identifying image features that matched the particle size. Particle coordinates were then linked between frames into time-dependent trajectories using trackpy,\cite{allan_soft-mattertrackpy_2024} which implements the Crocker-Grier algorithm.\cite{crocker_methods_1996} The resulting trajectories were converted to polar coordinates to extract the frame-to-frame displacement vectors $\delta \boldsymbol{q}$.\cite{ghosh_dynamics_2016} Displacement data from multiple trajectories were collected and sorted into radial bins according to the starting position of each vector. The spatially-varying drift and diffusion coefficients were then computed from the mean ($\mu_i$) and variance ($\sigma^2_{ij}$), respectively, of the binned displacement vectors $D^{(1)}_i = \mu_i / \Delta t$ and $D^{(2)}_{ij} = \sigma^2_{ij} / (2 \Delta t$), where $\Delta t$ is the imaging frame rate.

\subsection{Pattern writing}
Colloidal pattern writing was performed using the Pharos SEM controlled by an external computer running python scripts. Following a programmed sequence, patterns were drawn by successively raster scanning the electron beam in a square FOV centered on specified points on the droplet surface. These experiments used an accelerating voltage of 12 kV and 3.3 Hz imaging rate, with positive patterns using an electron flux of 2.0 $e^-/ \rm (nm^2 \cdot s)$ and negative patterns using a flux of 18.7 $e^-/ \rm (nm^2 \cdot s)$. For positive patterns, each point was scanned for 1.8 s (6 frames) to allow sufficient time for particle accumulation. For negative patterns, an exposure of 0.6 s (2 frames) was sufficient for particle depletion.

\subsection{Numerical modeling of solvent flow}
A 2D transport model was used to simulate the effects of electron-beam irradiation on solvent flow and particle transport, with flow velocity calculated according to the continuity equation. We assumed a uniform square sink distribution to correspond with the raster scanning pattern of the electron beam. The resulting velocity field was solved numerically in python. Full details of the model are provided in the Supporting Information.

\section*{Conflicts of Interest}
There are no conflicts to declare.

\section{Acknowledgment}

 We thank S. Aloni at the Molecular Foundry for assistance with the Zeiss SEM. This work has been supported by STROBE, A National Science Foundation Science \& Technology Center under Grant No. DMR 1548924. The SEM imaging at the Lawrence Berkeley Lab Molecular Foundry was performed as part of the Molecular Foundry user program, supported by the Office of Science, Office of Basic Energy Sciences, of the U.S. Department of Energy under Contract No. DE-AC02-05CH11231. N.S.G. acknowledges a David and Lucile Packard Foundation Fellowship for Science and Engineering.

\begin{suppinfo}

\begin{itemize}
    \item \textbf{Supporting Information}: Discussion of the pattern writing procedure; Analysis of the P\'{e}clet number; Voltage-dependence of particle diffusion; Analysis of the beam-particle interaction force with varying glycerol content; Discussion of the IL flow model
    \item \textbf{Video S1}: Low-voltage particle repulsion (4 kV, 0.37 $e^-/ \rm (nm^2 \cdot s)$, 500-nm-diameter particles)
    \item \textbf{Video S2}: Intermediate voltage particle dynamics (8 kV, 0.37 $e^-/ \rm (nm^2 \cdot s)$, 500-nm-diameter particles)
    \item \textbf{Video S3}: High-voltage particle attraction (12 kV, 0.37 $e^-/ \rm (nm^2 \cdot s)$, 500-nm-diameter particles)
    \item \textbf{Video S4}: Colloidal accumulation at low electron flux (12 kV, 2.0 $e^-/ \rm (nm^2 \cdot s)$, 500-nm-diameter particles)
    \item \textbf{Video S5}: Colloidal depletion at high electron flux (12 kV, 18.7 $e^-/ \rm (nm^2 \cdot s)$, 500-nm-diameter particles)
    \item \textbf{Video S6}: Dissipation of colloidal pattern writing
    \item \textbf{Video S7}: Attraction of a binary particle mixture (300- and 1000-nm-diameter particles)
\end{itemize}

\end{suppinfo}

\bibliography{maintext_bib.bib}

\end{document}